\documentclass[aps,amsmath,preprint,apl]{revtex4}
\usepackage[dvips]{graphicx}
\usepackage{bm}
\usepackage[dvips]{color}
\usepackage{setspace} 
\usepackage[dvips]{color} 

\begin{document}

\title{Generation of Spin Currents in the Skyrmion Phase of a Helimagnetic Insulator $\mathrm{Cu_2OSeO_3}$}
\author{Daichi Hirobe}%
\email{daichi.kinken@imr.tohoku.ac.jp}
\affiliation{Institute for Materials Research, Tohoku University, Sendai 980-8577, Japan}
\author{Yuki Shiomi}
\affiliation{Institute for Materials Research, Tohoku University, Sendai 980-8577, Japan}
\author{Yuhki Shimada}
\affiliation{Department of Physics, Toho University, 2-2-1, Miyama, Funabashi, Chiba 274-8510, Japan}
\author{Jun-ichiro Ohe}
\affiliation{Department of Physics, Toho University, 2-2-1, Miyama, Funabashi, Chiba 274-8510, Japan}
\author{Eiji Saitoh}
\affiliation{Institute for Materials Research, Tohoku University, Sendai 980-8577, Japan}
\affiliation{WPI Advanced Institute for Materials Research, Tohoku University, Sendai 980-8577, Japan}
\affiliation{CREST, Japan Science and Technology Agency, Chiyoda, Tokyo 102-0075, Japan}
\affiliation{The Advanced Science Research Center, Japan Atomic Energy Agency, Tokai 319-1195, Japan}
\date{\today}
\begin{abstract}
We report spin-current generation related with skyrmion dynamics resonantly excited by a microwave in a helimagnetic insulator $\mathrm{Cu_2OSeO_3}$. A Pt layer was fabricated on  $\mathrm{Cu_2OSeO_3}$ and voltage in the Pt layer was measured upon magnetic resonance of $\mathrm{Cu_2OSeO_3}$ to electrically detect injected spin currents via the inverse spin Hall effect (ISHE) in Pt. We found that ISHE-induced electromotive forces appear in the skyrmion phase of $\mathrm{Cu_2OSeO_3}$ as well as in the ferrimagnetic phase, which shows that magnetic skyrmions can contribute to the spin pumping effect.
\end{abstract}

\maketitle
\section{introduction}
A magnetic skyrmion, a nano-scale vortex-like spin texture in the real space, has attracted much attention in condensed matter physics.\cite{Nagaosa:2013,Fert:2013} Figure 1(a) illustrates a magnetic skyrmion. Magnetic moments point downward at the center of the skyrmion and upward along the periphery; intermediate magnetic moments vary their directions continuously such that the moments wrap the whole solid angle at the center. The skyrmion's diameter is typically $10$ to $100$ nm and the magnetic skyrmion has been usually observed in a crystalline form (the skyrmion crystal).\cite{Muhlbauer:2009,Yu:2010,Yu2:2010,Seki:2012} Since the skyrmion's spin configuration possesses the scalar spin chirality,\cite{Nagaosa:2013} topological phenomena appear in the Hall effect,\cite{Li:2013,Kanazawa:2011,Shiomi:2013,Neubauer:2009,Lee:2009,Schulz:2012} which is formulated by introducing an emergent electromagnetic field.\cite{Nagaosa:2013,Li:2013,Kanazawa:2011,Shiomi:2013,Neubauer:2009,Lee:2009,Schulz:2012}
Besides its rich physics, the skyrmion has potential for applications especially in spintronics. Unique properties of the skyrmion, i.e., particle-like nanostructure and topological stability, are highly promising for novel spintronic devices such as information storage or logic devices.\cite{Nagaosa:2013,Fert:2013} The skyrmion is also appealing as an efficient information carrier; in fact, recent experiments showed that magnetic skyrmions can be moved by electric current at low energy costs.\cite{Schulz:2012,Jonietz:2010,Yu:2012}
\par

In spite of the skyrmion's potential for spintronic applications, few studies have investigated skyrmions in terms of the pure spin current physics. A pure spin current refers to a flow of spin angular momentum with no charge currents; its understanding and control are central issues of spintronics.\cite{Zutic:2004} If the pure spin current can be generated from magnetic skyrmions, the generation should be interesting for basic research and would be a step forward in skyrmionic devices as well. Here, we experimentally demonstrate spin-current generation from a resonantly driven skyrmion motion in a helimagnetic insulator $\mathrm{Cu_2OSeO_3}$. Using a $\mathrm{Cu_2OSeO_3}$/Pt two-layer structure, we show that spin angular momentum transfers from the skyrmion crystal in $\mathrm{Cu_2OSeO_3}$ to conduction electrons in Pt via the spin pumping effect.\cite{Mizukami:2002,Tserkovnyak:2005,Azevedo:2005} The injected spin current is electrically detected in Pt via the inverse spin Hall effect (ISHE),\cite{Kimura:2007,Saitoh:2006,Valenzuela:2006,Seki:2008,Takahashi:2002,Ando:2008} which converts a spin current into a transverse electromotive force $\mathbf{E}_{\mathrm{SHE}}$ via large spin-orbit interaction of Pt [Fig. 1(b)]. $\mathbf{E}_{\mathrm{SHE}}$ is given by\cite{Saitoh:2006}
\begin{equation}
\mathbf{E}_{\mathrm{SHE}} \propto \mathbf{J}_{\mathrm{s}}\times\mathbf{\boldsymbol\sigma}, \label{ishe}
\end{equation}
when a spin current carries the spin polarization $\mathbf{\boldsymbol\sigma}$ in the spatial direction $\mathbf{J}_{\mathrm{s}}$.
\par

We used $\mathrm{Cu_2OSeO_3}$ for this work because the high-insulating property of $\mathrm{Cu_2OSeO_3}$ enables electric detection of spin currents free from galvanomagnetic contamination.\cite{Chen:2013} The skyrmion crystal in $\mathrm{Cu_2OSeO_3}$ has been observed by Lorentz transmission electron microscopy\cite{Seki:2012} and neutron diffraction measurements.\cite{Adams:2012,Seki2:2012} The helimagnetic ordering temperature is approximately $60$ K and a helical spin structure is stabilized by the Dzyaloshinskii-Moriya interaction at the zero magnetic field. The skyrmion-crystal phase appears in the vicinity of the ordering temperature in a restricted window of magnetic fields for bulk $\mathrm{Cu_2OSeO_3}$ crystals.\cite{Seki:2012} Skyrmions form a two-dimensional triangular lattice within the plane perpendicular to an applied magnetic field.\cite{Seki:2012,Adams:2012,Seki2:2012} The magnetic phase diagram of $\mathrm{Cu_2OSeO_3}$ is almost the same for any direction of an applied magnetic field.\cite{Adams:2012} 
\par

\section{method}
We grew single crystals of $\mathrm{Cu_2OSeO_3}$ by a chemical vapor transport method.\cite{Miller:2010} The volume of the crystals was typically $1$ mm$^{3}$. The magnetic moment $M$ for the prepared crystals was investigated with a vibrating sample magnetometer (VSM) in a Physical Properties Measurement System (Quantum Design, Inc.). To make a $\mathrm{Cu_2OSeO_3}$/Pt sample, we polished the surface of $\mathrm{Cu_2OSeO_3}$ and sputtered a $5$-nm-thick Pt film on the polished surface in an argon atmosphere. Figure 1(c) illustrates an experimental set-up for detecting the spin pumping effect. To conduct microwave experiments, we used a reflection-type coplanar waveguide and a network analyzer. Since the skyrmion phase is restricted in a narrow window of magnetic fields, we fixed a magnetic field and swept a microwave frequency to measure magnetic resonance. A static in-plane magnetic field was applied perpendicularly to a microwave magnetic field. When a microwave frequency fulfills resonance conditions of $\mathrm{Cu_2OSeO_3}$, magnetic moments are forced to precess (magnetic resonance). If this precession transfers spin angular momentum to conduction electrons in Pt, the spin current is converted into $\mathbf{E}_{\mathrm{SHE}}$ in Pt via the ISHE. To measure $\mathbf{E}_{\mathrm{SHE}}$, we attached two electrodes to the ends of the Pt film [Fig. 1(c)]. Applied microwaves were set at 10 mW. The used $\mathrm{Cu_2OSeO_3}$ sample was of column shape and 1.5 mm across and 0.5 mm high. The spin-pumping measurement was performed in a cryogenic probe station, where the sample was cooled by attaching the bottom of a sample holder to a cold head. A system temperature was measured with a thermometer near the cold head; the sample temperature appeared higher than the system temperature by about 5 K in the present study. 

\section{results and discussion}
The prepared $\mathrm{Cu_2OSeO_3}$ single crystals were characterized by measuring $M$. Figure 2 shows temperature dependence of $M$ at 300 mT. With decreasing temperature, $M$ sharply increases at approximately 60 K and saturates at $0.52$ $\mu_{\mathrm{B}}/\mathrm{Cu}^{2+}$ at 2 K. The ordering temperature and the saturation moment are consistent with those of preceding studies.\cite{Bos:2008} The inset to Fig. 2 shows magnetic-field dependence of $M$ at 5 K. $M$ saturates above 100 mT, which suggests transition into a ferrimagnetic phase. In the magnetization curve below 100 mT, a step-like change appears near 30 mT. This change corresponds to a phase transition between multi-domain and single-domain helimagnetic states.\cite{Seki:2012,Bos:2008} Hence, the magnetic properties of the prepared crystals are well consistent with the reported ones.

Figure 3 shows microwave responses of a $\mathrm{Cu_2OSeO_3}$/Pt sample at 45 K and 50 K. Hereafter, temperatures refer to a system temperature $T$ of the probe station. Figure 3(a) shows frequency dependence of microwave absorption ${\it{\Delta}} S_{11}$ at 45 K, which is well lower than the helimagnetic ordering temperature. At 114 mT, ${\it{\Delta}} S_{11}$ shows a dip around 3.5 GHz. Since the resonance frequency decreases with decreasing a magnetic field, the microwave absorption is attributed to a ferrimagnetic resonance of $\mathrm{Cu_2OSeO_3}$. Below 54 mT, two dips appear and their resonance frequencies increase with decreasing a magnetic field. The field dependence of the resonance frequencies is attributed to a helimagnetic phase induced by the Dzyaloshinskii-Moriya interaction.\cite{Kataoka:1987, Onose:2012} Note that the two dips in the helimagnetic phase correspond to two spin-wave modes.\cite{Kataoka:1987,Onose:2012,Date:1977} Figure 3(c) summarizes field dependence of resonance frequencies at 45 K.

We observed magnetic resonances of the skyrmion crystal between 47 K and 51 K.  Figure 3(b) shows frequency dependence of ${\it{\Delta}} S_{11}$ at 50 K. The resonance frequency decreases with decreasing a magnetic field from $84$ mT, and then increases below $39$ mT. This behavior is similar to that of $45$ K. From 30 mT through 6 mT, however, an additional excitation appears around 1.4 GHz while the excitation around 2 GHz is suppressed. Below 3 mT, the additional dip disappears and the helimagnetic one revives around 2 GHz. Figure 3(d) summarizes field dependence of resonance frequencies at 50 K. While helimagnetic and ferrimagnetic modes appear both at 45 K and at 50 K, the new resonance mode appears at $50$ K only, as highlighted by the shading of Fig. 3(d). Its resonance frequency is lower than those of the other modes and monotonically increases with increasing a magnetic field. This magnetic-field dependence obviously differs from that of helimagnetic modes. Moreover, the new magnetic excitation occurs in a restricted window of magnetic fields. These results strongly indicate that magnetic excitations near 1.5 GHz originate from a collective motion of the skyrmion crystal. According to preceding studies on magnetic resonance of skyrmion crystals,\cite{Onose:2012,Petrova:2011,Mochizuki:2012} the newly observed mode is attributed to a counterclockwise rotation of magnetic skyrmions, in which skyrmion's core rotates counterclockwise. Note that the excitation observed around 2 GHz in the skyrmion-crystal phase is due to either a clockwise rotation of skyrmions\cite{Okamura:2013} or magnetic resonance of remnant helimagnetic states\cite{Muhlbauer:2009,Onose:2012}; the origin of this excitation was unidentified in this experiment. Accordingly, we confine ourselves to the counterclockwise skyrmion motion to investigate the spin pumping effect.

Figure 4(a) shows a skyrmion phase diagram determined by magnetic resonance in the $T$ range slightly below the ordering temperature. Here, critical magnetic fields $H_{\mathrm{c}1}$, $H_{\mathrm{c}2}$, and $H_{\mathrm{c}3}$ are defined as shown in Fig. 3(d); from $H_{\mathrm{c}1}$ through $H_{\mathrm{c}2}$ the skyrmion excitation continues while across $H_{\mathrm{c}3}$ a phase transition occurs from the helimagnetic state to the ferrimagnetic state. $T$ dependence of the critical magnetic fields reveals that magnetic excitations of the skyrmion crystal are restricted in a narrow window of $T$ (approximately 4 K). This observation is consistent with a phase diagram determined by measurements of ac magnetic susceptibility,\cite{Seki:2012,Onose:2012} although error bars for $H_{\mathrm{c}1}$ and $H_{\mathrm{c}2}$ are rather large because of small skyrmion excitations around $H_{\mathrm{c}1}$ and $H_{\mathrm{c}2}$. Note again that the skyrmion-crystal phase was observed at $T$ lower than those reported elsewhere\cite{Seki:2012,Adams:2012,Seki2:2012} because in the present set-up we could not attach the thermometer directly to the sample as mentioned in the method part.

We demonstrate spin-current generation by measuring ISHE-induced electromotive forces in Pt at $50$ K. We first investigated the ferrimagnetic spin pumping at 66 mT [point A in Fig. 4(a)]. Figures 4(b) and 4(c) represent a microwave-absorption spectrum and an electromotive-force spectrum at point A, respectively. In the microwave-absorption spectrum [Fig. 4(b)], a single dip of ferrimagnetic resonance appears around 2 GHz. Correlating with the resonance, a single dip of electromotive forces appear around the resonance frequency, as shown in Fig. 4(c). Moreover, the electromotive forces show opposite polarity with an opposite magnetic field. This reversal is consistent with the ISHE [Eq. (\ref{ishe})] and provides evidence for spin-current generation from the ferrimagnetic resonance.

Next, we conducted the spin-pumping measurement in the skyrmion-crystal phase [point B in Fig. 4(a)]. A magnetic field was set at 15 mT to maximize microwave-absorption intensity within the skyrmion-crystal phase [see also Fig. 3(b)]. Figures 4(d) and 4(e) are a microwave-absorption spectrum and an electromotive-force spectrum at point B, respectively. In the microwave-absorption spectrum [Fig. 4(d)], two absorption dips appear; the dip at the lower frequency ($\approx$1.3 GHz) corresponds to the counterclockwise rotation of skyrmions [see also Figs. 3(b) and 3(d)]. In the electromotive-force spectrum [Fig. 4(e)], electromotive forces appear around 1.3 GHz in correlation with the skyrmion excitation. Again, the electromotive forces show opposite polarity with an opposite magnetic field. Moreover, the electromotive force is of the same sign for skyrmion and ferrimagnetic phases for a given field direction. The results are consistent with Eq. (\ref{ishe}) because the direction of spin polarization should be determined by a static magnetic field also in the skyrmion-crystal phase. These observations show that the resonantly excited skyrmion motion can give rise to the spin pumping effect and that the skyrmion dynamics can be detected electrically via the ISHE. Dividing the voltage by the intensity of microwave absorption\cite{Iguchi:2012} yields 16 $\mathrm{\mu V/W}$ for the ferrimagnetic phase and 8 $\mathrm{\mu V/W}$ for the skyrmion-crystal phase, which implies the lower efficiency of the spin pumping in the skyrmion-crystal phase.

We numerically calculated spin currents related with skyrmion dynamics to demonstrate theoretically that the counterclockwise skyrmion motion can contribute to the spin pumping effect. The dynamics of local magnetizations $\mathbf{M}(\mathbf{r},t)$ were simulated with the Landau-Lifshitz-Gilbert (LLG) equation\cite{Mochizuki:2012,Ohe:2013} for a single skyrmion on a two-dimensional square lattice in the $x$-$y$ plane. The used Hamiltonian is
\begin{equation}
\mathcal{H} = -J\sum_{i,j}\mathbf{M}_{i}\cdot\mathbf{M}_{j}+D\sum_{i}\left(\mathbf{\hat{x}}\cdot\left(\mathbf{M}_{i}\times\mathbf{M}_{i+\mathbf{\hat{x}}}\right) + \mathbf{\hat{y}}\cdot\left(\mathbf{M}_{i}\times\mathbf{M}_{i+\mathbf{\hat{y}}}\right)\right)-\sum_{i}\mathbf{M}_{i}\cdot\mathbf{H},
\end{equation}
where $J=3.9$ meV and $D=0.5J$ are coupling constants of ferromagnetic interaction and Dzyaloshinskii-Moriya interaction, respectively, and $\mathbf{H}=H\mathbf{\hat{z}}$ is a static magnetic field in the $z$ direction, i.e., perpendicular to the square lattice. The system size was $41 \times 41$ sites. The parameters used for the LLG equation were: the gyromagnetic ratio $\gamma = 1.76\times10^{11}$ Hz/T; the Gilbert damping constant $\alpha = 0.01$; a static magnetic field $\mu_0H=800$ mT in the $z$ direction; a microwave magnetic field $\mu_0h =  8$ mT in the $y$ direction; the microwave frequency $f=0.78$ GHz. With these parameters, the skyrmion's core resonantly rotates counterclockwise as shown in Figs. 5(a)-(d). Note that values of the static magnetic field and the resonance frequency are not equal to those of our measurement because the simplest model of a single skyrmion was adopted to capture the essence of the spin pumping from the skyrmion dynamics.

Generated spin currents were calculated using the time average of $D_z(t) = \left(\mathbf{M}\times\partial\mathbf{M}/\partial t\right)_{z}$\cite{Ando:2011} at the bottom edge of the system $\mathbf{r}=(x,0)$, namely $\left<{J_\mathrm{s}}^z\right> \propto \int\mathrm{d}x\left<D_z(t)\right>_{\mathrm{time}}$, where $\left< D_z(t) \right>_{\mathrm{time}}$ represents the time average of $D_z(t)$. As shown in Figs. 5(e)-(h), both the magnitude and the sign of $D_z(t)$ depend heavily on space and time. The spatial dependence reflects a swirling spin structure due to the Dzyaloshinskii-Moriya interaction; such behavior has not been observed for centrosymmetric systems such as $\mathrm{Y_3Fe_5O_{12}}$, where the Dzyaloshinskii-Moriya interaction is absent. The spatial and temporal dependence of $D_z(t)$ at the edge gives rise to the complicated time evolution of generated spin currents $J_s^z(t)$ as indicated by a blue dashed line in Fig. 5(i). The temporal dependence of $J_s^z(t)$ cannot be described by a single sine wave, which implies that nonlinear magnetization dynamics were also driven for $\mu_0h=8$ mT. Nevertheless, $\left<{J_\mathrm{s}}^z\right>$ is found to be finite as highlighted by the shading in Fig. 5(i), which theoretically substantiates our experimental results. As illustrated in Figs. 5(a)-(d), magnetic moments even at the edge form a noncollinear configuration owing to the Dzyaloshinskii-Moriya interaction. Although a similar noncollinear configuration at the edge was found also for the ferrimagnetic phase in the calculation with $\mu_0 H=2.4$ T, the temporal dependence of $J_s^z(t)$ for $\mu_0h=8$ mT was simply sinusoidal. We note that $\left<{J_\mathrm{s}}^z\right>$ for the skyrmion phase was approximately 30 \% smaller than for the ferrimagnetic phase in the present calculation. This result indicates the lower efficiency of the spin pumping in the skyrmion phase, which is consistent with our experimental result. The numerical calculation presented here demonstrates that non-uniform spin configurations such as skyrmions can give rise to the spin pumping effect while the observation of the spin pumping has been confined to uniform spin configurations such as ferromagnetic and ferrimagnetic states so far.

\section{summary}
To summarize, we have observed spin-current generation from skyrmion dynamics in a $\mathrm{Cu_2OSeO_3}$/Pt two-layer structure. We resonantly excited magnetic skyrmions in $\mathrm{Cu_2OSeO_3}$ by a microwave and electrically detected spin currents via the ISHE in Pt. A numerical calculation showed that a rotational motion of skyrmions can give rise to a finite dc spin current via the spin pumping effect. Spin-current generation via skyrmion dynamics demonstrated here can be a new functionality operating at microwave frequencies in spintronic applications.

We thank S. Seki for fruitful discussions and R. Iguchi for technical assistance. This work was supported by CREST-JST ``Creation of Nanosystems with Novel Functions through Process Integration," Strategic International Cooperative Program ASPIMATT from JST, and Grants-in-Aid for Scientific Research (A) from JSPS (No. 24244051) and for Scientific Research on Innovative Areas from MEXT (``Topological Quantum Phenomena'' (No. 25103702)) as well as by MEXT KAKENHI Grant Number 23108004.

\newpage{}

\newpage{}
\begin{figure}[htbp]
  \includegraphics[width=12.0cm]{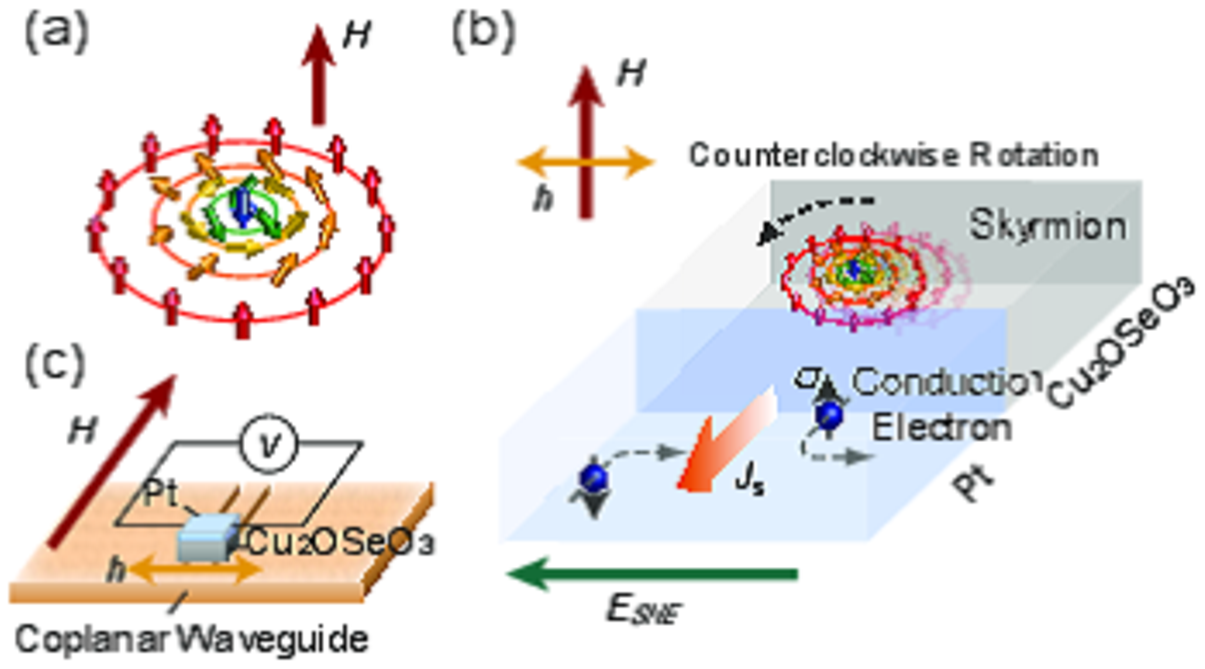}
\caption{(a) A schematic illustration of a magnetic skyrmion. Arrows represent magnetic moments. (b) A schematic illustration of the inverse spin Hall effect induced by the skyrmion spin pumping in a $\mathrm{Cu_2OSeO_3}$/Pt two-layer structure. A microwave magnetic field $h$ is applied perpendicularly to a static magnetic field $H$. $\sigma$ and $J_{\mathrm{s}}$ denote the spin polarization and the spatial direction of a spin current, respectively. $E_{\mathrm{SHE}}$ denotes an electromotive force induced by the inverse spin Hall effect. (c) Experimental set-up for the spin-pumping measurement.}  
  \label{Fig.1.}
\end{figure}

\newpage{}
\begin{figure}[htbp]
  \includegraphics[width=6.0cm]{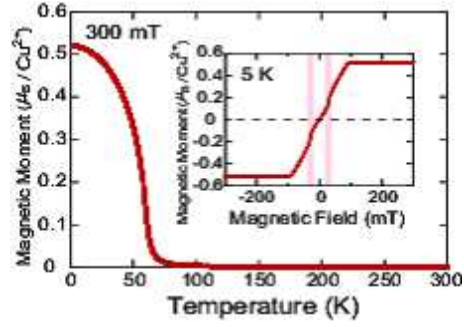}
\caption{Temperature dependence of the magnetization of $\mathrm{Cu_2OSeO_3}$. An applied magnetic field was set at $300$ mT. The inset shows magnetic-field dependence of the magnetization of $\mathrm{Cu_2OSeO_3}$ at 5 K. The shading at $\pm30$ mT highlights transitions between multi-domain and single-domain helimagnetic phases.}  
  \label{Fig.2.}
\end{figure}

\begin{figure}[htbp]
  \includegraphics[width=12.0cm]{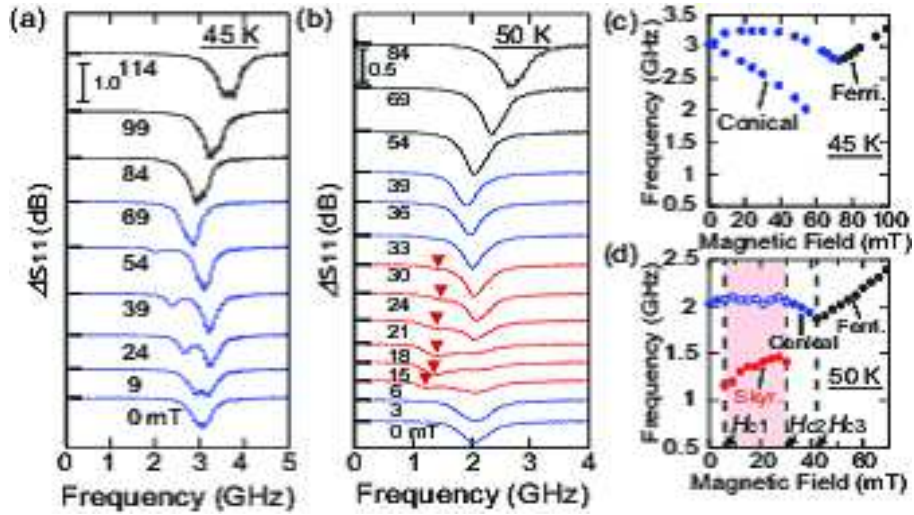}
    \caption{(a),(b) Frequency dependence of microwave absorption (a) at $45$ K and (b) at $50$ K at various magnetic fields. Black and blue curves represent data in ferrimangetic and helimagnetic phases, respectively. Red triangles in (b) indicate the magnetic resonance of the skyrmion crystal. Microwave-absorption spectra are shifted vertically for clarity. (c),(d) Magnetic-field dependence of resonance frequencies (c) at $45$ K and (d) at $50$ K. The skyrmion resonance persists from the critical field $H_{\mathrm{c}1}$ through the critical field $H_{\mathrm{c}2}$. Across the critical field $H_{\mathrm{c}3}$, a transition occurs between helimagnetic and ferrimagnetic phases.}
  \label{Fig.3.}
\end{figure}

\begin{figure}[htbp]
  \includegraphics[width=12.0cm]{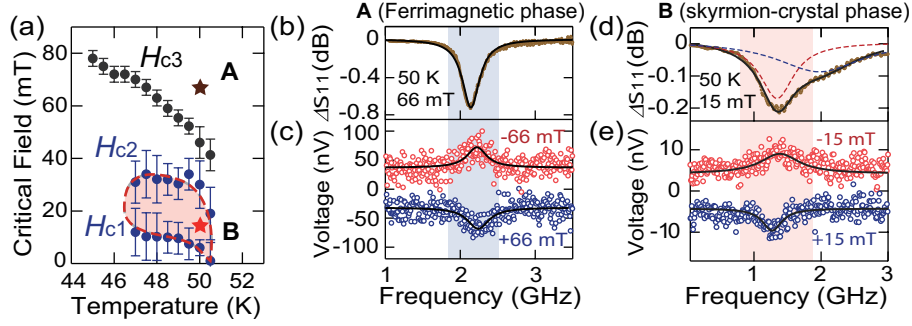}
    \caption{(a) Temperature dependence of critical magnetic fields $H_{\mathrm{c}1}$, $H_{\mathrm{c}2}$, and $H_{\mathrm{c}3}$. For definitions of $H_{\mathrm{c}1}$, $H_{\mathrm{c}2}$, and $H_{\mathrm{c}3}$, see the main text and Fig. 3(d). The light pink background shows the skyrmion-crystal phase. Spin-pumping measurements were conducted at points A (ferrimagnetic phase) and B (skyrmion-crystal phase). (b)-(e) Frequency dependence of (b),(d) microwave absorption and (c),(e) electromotive forces at 50 K. (b) and (c) are for the ferrimagnetic phase at 66 mT; (d) and (e) are for the skyrmion-crystal phase at 15 mT. Symbols are experimental data and solid curves are (multi-)Lorentzian fits to the data. A red dashed curve in (d) represents skyrmion-crystal resonance. }
  \label{Fig.4.}
\end{figure}

\begin{figure}[htbp]
  \includegraphics[width=12.0cm]{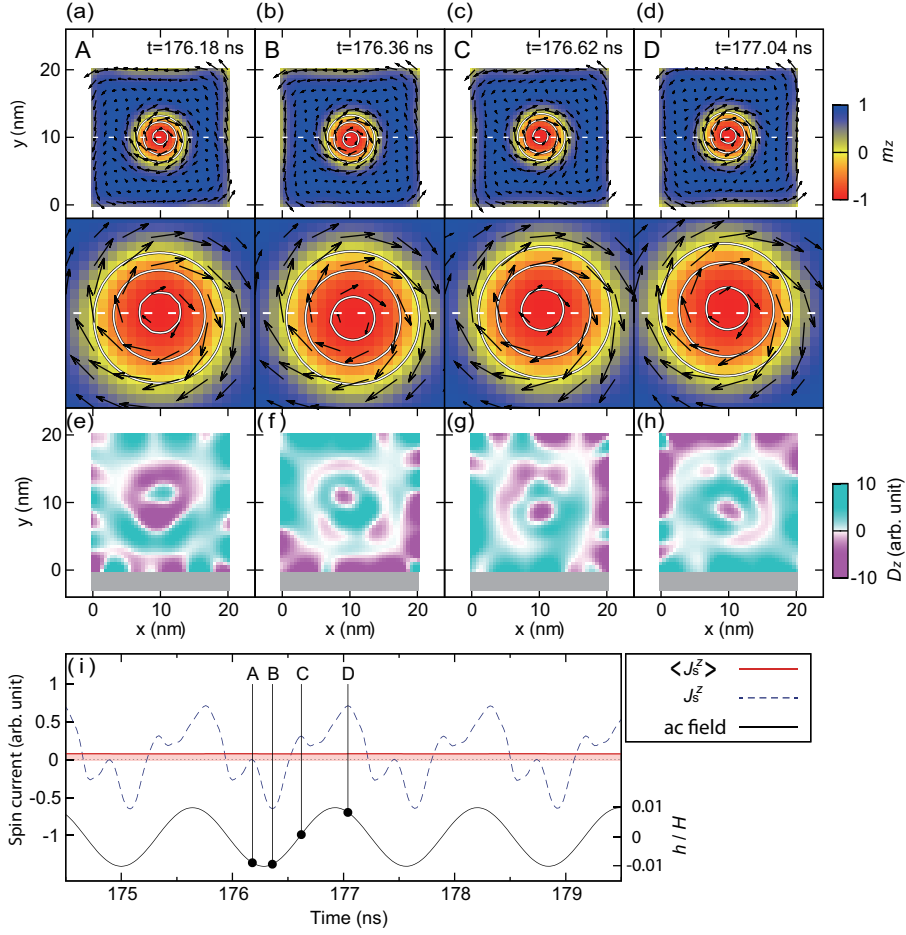}
    \caption{(a)-(d) Representative snapshots of the couterclockwise skymion rotation. Black arrows represent $x$ and $y$ components of local magnetizataions while the contour plot represents the $z$ component. Magnified views are shown in the panels below (a)-(d). (e)-(h) Representative snapshots of the spatial distribution of the damping torque, $D_z(t) = \left(\mathbf{M}\times\partial\mathbf{M}/\partial t\right)_{z}$ for the skyrmion phase. Spin currents generated at the edge, $y=0$, were calculated. (i) Time evolution of generated spin currents ${J_\mathrm{s}}^z$ for the skyrmion phase (blue dashed curve) and a microwave magnetic field divided by a static magnetic field $h/H$ (black solid curve). The red line represents the magnitude of a time-averaged spin current $\left<{J_\mathrm{s}}^z\right>$.}
  \label{Fig.5.}
\end{figure}

\end{document}